\documentclass[letterpaper]{article}

\usepackage[T1]{fontenc}

\usepackage{geometry}
\usepackage{setspace}

\usepackage{achemso}

\usepackage{graphicx}
\usepackage{float}
\newfloat{scheme}{htbp}{los}
\floatname{scheme}{Scheme}
\floatname{chart}{Chart}
\newfloat{graph}{htbp}{loh}

\usepackage{chemformula} 
\usepackage[version = 4]{mhchem} 

\usepackage{authblk}
\author[1]{O.~S.~Ken*}
\author[1]{D.~O.~Horiachyi}
\author[1]{V.~L.~Korenev}
\author[1]{A.~V.~Trifonov}
\author[1]{I.~A.~Akimov}
\author[1]{D.~R.~Yakovlev}
\author[1, 2]{M.~Bayer}
\affil[1]{Experimentelle Physik 2, Technische Universit\"at Dortmund, 44221 Dortmund, Germany}
\affil[2]{Research Center FEMS, Technische Universit\"{a}t Dortmund, 44227 Dortmund, Germany}

\title{Bragg-enhanced time-domain Brillouin scattering from a propagating acoustic grating}
\date{*Email: olga.ken@tu-dortmund.de}

\begin{document}

\maketitle

\begin{abstract}
  Generation and detection of coherent phonons in semiconductors by femtosecond optical pulses is a powerful tool for high-frequency acoustic control of their electronic properties. Here, we demonstrate a propagating one-dimensional acoustic grating in bulk semiconductors using above-band-gap excitation by a train of laser pulses with high repetition rate of 1~GHz. This approach enables shaping of the coherent acoustic phonon spectrum and leads to a significant enhancement of Brillouin light scattering at selected probe wavelengths in pump-probe configuration. We demonstrate this effect at a cryogenic temperature of 5~K in prototypical semiconductor systems, namely bulk crystalline GaAs and (Cd,Zn)Te, which serve as benchmark materials for the proposed method. The spectral dependence of the time-domain Brillouin light scattering amplitude exhibits resonant peaks at discrete probe wavelengths arising from Bragg reflection of the probe light by the propagating acoustic grating. A $10-30$-fold resonant enhancement of the signal amplitude is observed for GaAs and (Cd,Zn)Te, determined by the finite spectral width of the probe pulse. With further spectral narrowing, enhancements of the order $\sim 100-150$ are expected, set by the number $N$ of strain pulses in the acoustic grating within the sample and ultimately limited by the material parameters and sample thickness.
\end{abstract}

\section*{Keywords}

time-resolved Brillouin scattering, coherent acoustic phonons, acoustic grating, Bragg grating, ultrafast photoacoustics

\section{Introduction}

Ultrafast laser photoacoustics~\cite{Thomsen1986, Wright1994} provides a versatile approach for probing lattice dynamics in a wide range of materials~\cite{Matsuda2015, Gusev2018, Ruello2015, Audoin2023}. Absorption of a femtosecond optical pulse in the near-surface region launches (through the optoacoustic effect) a strain pulse that propagates into the material and can be considered as a coherent acoustic phonon wave packet~\cite{Gusev2018, Matsuda2015}. The generation of coherent phonons in semiconductors exploits excitation of ultrafast stress associated with the electronic deformation potential in semiconductors, i.e. the expansion of the lattice due to photogenerated charge carriers (electrons and holes)~\cite{Thomsen1986,Gusev2012,Fainstein2012}. In pump–probe experiments, detection is typically based on transient changes in the reflectivity of a time-delayed probe pulse. Wave packets of coherent phonons modulate the optical properties of the material; however, their direct coupling to light is typically weak. To enhance detection sensitivity in transparent materials, heterodyne detection is commonly employed, in which weak probe light pulse scattered by coherent acoustic phonons interferes with a much stronger probe field reflected from the sample surface~\cite{Thomsen1986, Gusev2018}. This interference gives rise to an oscillatory component in the differential reflectivity signal measured in pump-probe as a function of time delay. This interferometric technique is known as time-domain Brillouin light scattering (TDBS)~\cite{Ruello2015, Matsuda2015, Gusev2018} and can be used for nanoimaging~\cite{Audoin2023} and high sensitivity measurements of attometer vibrations~\cite{Karzel2025}. Moreover, high frequency coherent acoustic phonons are appealing for sub-THz control of the optical response in semiconductors~\cite{Fainstein2024,Bruggemann2012,Baldini2019}.

The physical principles underlying TDBS are closely related to those of transient grating~\cite{TrGrat1, TrGrat2} and impulsive stimulated Brillouin scattering~\cite{StimulatedBS1, Nelson1987} techniques, in which two coherent pump beams intersect to form a spatially periodic excitation pattern. Such transient optical gratings can launch coherent acoustic phonons via electrostriction, thermal expansion, or deformation-potential coupling, resulting in a time-dependent strain field with standing-wave character. The resulting spatially periodic strain modulation acts, through the associated variation of the dielectric function, as a dynamic diffraction grating that coherently scatters a time-delayed probe beam incident at the Bragg phase-matching condition.

\begin{figure}
  \centering
  \includegraphics[width=0.9\linewidth]{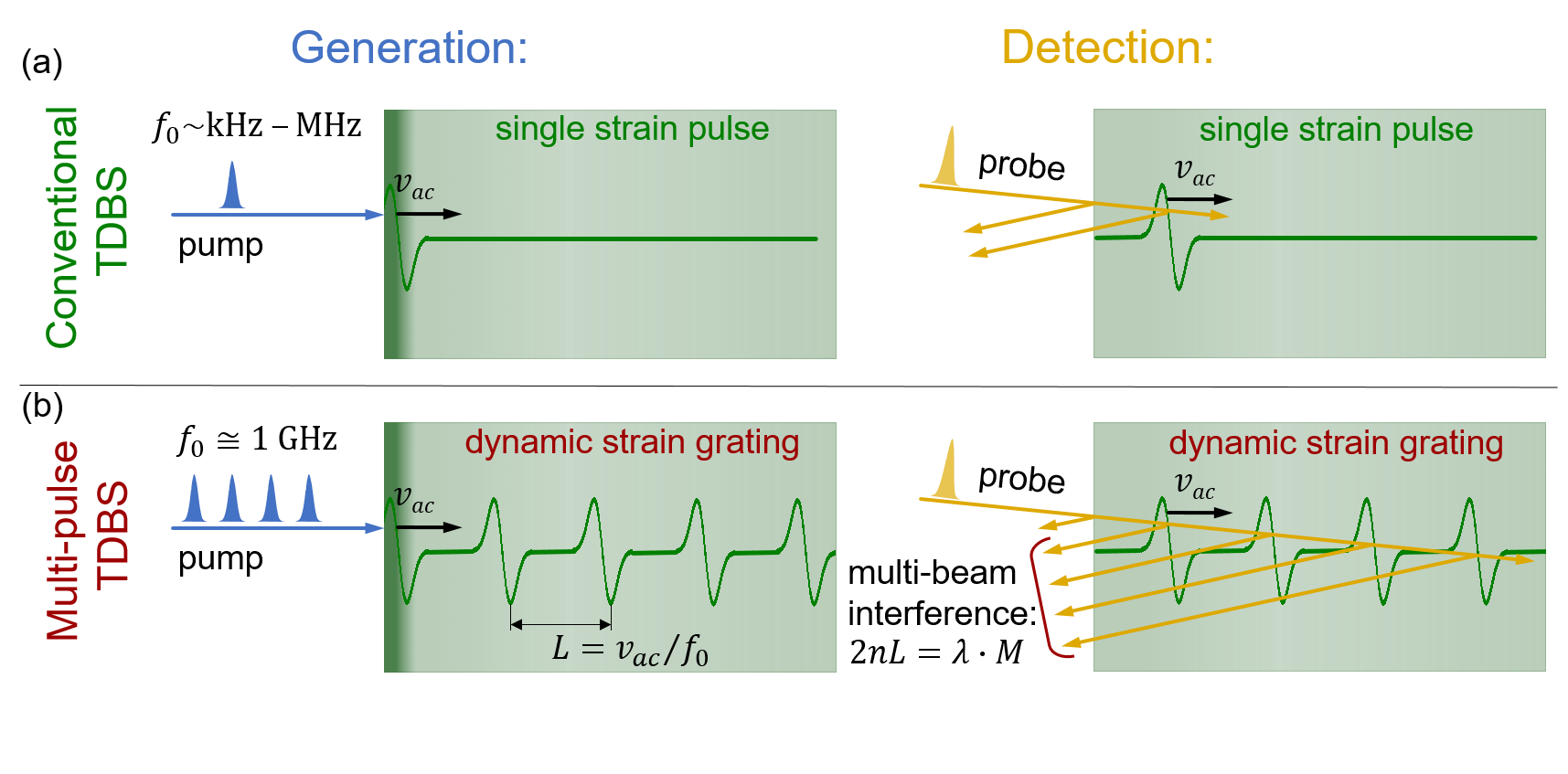}
  \caption{Schematic illustration of the conventional and multi-pulse TDBS approaches. (a) In conventional TDBS, a single pump pulse generates a propagating strain pulse that scatters the probe light; interference between the scattered field and the probe light reflected from the sample surface gives rise to the TDBS signal. (b) In the multi-pulse approach, a high-repetition-rate ($\sim1\,\mathrm{GHz}$) train of pump pulses generates a propagating acoustic grating. When the Bragg condition is satisfied, constructive interference of probe light scattered from successive grating planes leads to a pronounced enhancement of the TDBS signal amplitude.}
  \label{fig:Generation}
\end{figure}

Recent advances in semiconductor lasers with repetition rates up to hundreds of GHz, enabling compact and monolithically integrated implementations~\cite{Chang2022,Buyalo2024}, offer new opportunities for ultrafast photoacoustics. When a system is driven by a periodic train of ultrashort pulses with a repetition rate approaching the relevant vibrational frequencies (hundreds of MHz to GHz), resonant buildup of the  oscillation amplitude can occur.
Such effects, analogous to resonance phenomena in optical cavities, were demonstrated for both phonons \cite{Nelson2005, Bojahr2013, Nelson2011, SubharmCombMembr2011} and magnons \cite{Aleman2020, Jackl2017, Kobecki2020}, enabling frequency-selective excitation of vibration modes. 
However, to the best of our knowledge, they have not yet been investigated in bulk semiconductor systems, where successive coherent strain pulses generated at GHz repetition rates can form a dynamic acoustic grating within the three-dimensional volume of the material, giving rise to an enhanced optical response (such as increased reflectivity or Brillouin scattering of the probe light) under Bragg-matching conditions.
 
In this work, we demonstrate an approach that extends conventional TDBS by exploiting a dynamic acoustic Bragg grating generated by a periodic train of pump pulses with 1~GHz repetition rate. Under such excitation, coherent acoustic strain pulses launched from the surface build up a spatially periodic strain distribution that propagates into the bulk at the sound velocity.  
In the proposed approach, the acoustic grating (i) is generated solely by a single laser beam, (ii) by repeated absorption of optical pulses near the surface of a semiconductor, and (iii) is formed sequentially in time.
The proposed approach is validated on prototypical semiconductor systems, namely bulk crystalline GaAs and CdZnTe. We provide a quantitative analysis of the interference mechanism and its spectral selectivity, which accounts for the experimental observations. Constructive interference of the probe light fields scattered from the acoustic grating leads to a $10-30$-fold resonant enhancement of the TDBS signal at discrete probe wavelengths satisfying the Bragg condition.  We show that the enhancement factor relative to conventional single-pulse TDBS is governed by acoustic attenuation and optical absorption coefficients of the material. In the limit of weak attenuation, the enhancement factor can reach values $\sim 100-150$ upon further spectral narrowing of the probe, with the ultimate limit set by the number $N$ of strain pulses forming the acoustic grating within the sample volume.

\section{Methods}

\subsection{TDBS under excitation with high repetition rate}

The proposed multi-pulse approach is schematically illustrated in Fig.~\ref{fig:Generation}, together with a comparison to conventional single-pulse TDBS. In the conventional scheme (Fig.~\ref{fig:Generation}(a)), absorption of a femtosecond pump pulse in the near-surface region of the sample induces an ultrafast expansion of the crystal lattice, launching a coherent acoustic strain pulse that propagates into the bulk with the sound velocity $v_{\mathrm{ac}}$. As the strain pulse travels through the material, it modulates the local dielectric permittivity. For probe wavelengths within the transparency region, the probe light penetrates into the sample and is partially scattered by the propagating strain pulse. Interference between this scattered field and the probe light reflected from the sample surface gives rise to oscillations in the transient reflectivity signal, $\Delta R(t)/R$, with frequency 
\begin{equation}
f_{\mathrm{ac}}=2v_{\mathrm{ac}}n/\lambda=2v_{\mathrm{ac}}n\nu/c,
\end{equation}
where $\nu$ is the frequency of the probe light, $\lambda = c/\nu$ is its wavelength and $c$ is the speed of light in vacuum, $n$ is the refractive index.

In contrast, under high-repetition-rate excitation (Fig.~\ref{fig:Generation}(b)), a periodic train of pump pulses generates a sequence of strain pulses that propagate into the bulk. The separation between successive pulses is $L = v_{\mathrm{ac}}/f_{0}$, which, for the laser repetition frequency $f_{0}\approx 1\, \mathrm{GHz}$ and typical sound velocities $v_{\mathrm{ac}}$ of several km/s, is on the order of micrometers. This periodic strain distribution forms a propagating acoustic Bragg grating. Constructive interference of probe light scattered from this grating occurs only at specific probe wavelengths, $\lambda_M$, satisfying the Bragg condition:

\begin{equation}
M\lambda_M/n=2L,
\label{eq: resonance}
\end{equation}
where $M$ is an integer and we consider normal incidence of the probe beam. Under these conditions, interference of the probe light scattered from successive strain pulses leads to a pronounced enhancement of the amplitude of the $\Delta R(t)/R$ signal at particular probe wavelength.

This process can also be viewed as a modification of the nonequilibrium acoustic phonon spectrum under high-repetition-rate excitation. In the single-pulse regime, the strain pulse is characterized by a broad frequency spectrum. In contrast, excitation by a periodic train of pump pulses leads to spectral narrowing~\cite{Nelson2005, Nelson2011, Aleman2020}, with pronounced peaks emerging at frequencies that are integer multiples of the repetition rate. Accordingly, only nonequilibrium phonons satisfying $f_{\mathrm{ac}} = 2 v_{\mathrm{ac}} n / \lambda_M = M f_0$ are efficiently generated.

\begin{table}
  \caption{Energy of the exciton optical transition, $E_{\mathrm{X}}$, at a temperature of $T \approx 5\, \mathrm{K}$, absorption coefficient $\alpha_{\mathrm{pump}}$ of the pump light, refractive index $n$ in the transparency range of probe wavelengths, sound velocity $v_{\mathrm{ac}}$ along [100] and acoustic grating period $L$ for GaAs and (Cd,Zn)Te. Literature values are cited in the table; parameters without citation are obtained in the present work. }
  \label{tbl:Parameters}
  \centering
  \begin{tabular}{lccccc}
    \hline
    & $E_{\mathrm{X}}$ (eV) & $\alpha_{\mathrm{pump}}$ ($10^{5}\,\mathrm{cm}^{-1}$) & $n$ & $v_{\mathrm{ac}}$ (km/s) &  $L\, (\mathrm{\mu m)}$\\
    \hline
   GaAs & 1.515~\cite{Sell1973} & 6~\cite{Sell1975} & 3.59--3.66 & 4.73~\cite{Blakemore1982} & 4.74 \\
    Cd$_{0.88}$Zn$_{0.12}$Te & 1.664~\cite{Godde2013} & 2~\cite{OptConstCdZnTe} & 2.86--2.96~\cite{CdZnTe_refr} & 3.15 & 3.16 \\
    \hline
  \end{tabular}
\end{table}

\subsection{Experimental details}

The proposed experimental approach was applied to prototypical semiconductor systems, namely single-crystalline bulk [100] GaAs and [100] Cd$_{0.88}$Zn$_{0.12}$Te. The parameters of GaAs are well established in the literature, while details of the growth and optical properties of Cd$_{0.88}$Zn$_{0.12}$Te are given in Ref.~\cite{GoddeAkimov2010}. The key material parameters used in the analysis and for the estimates presented below are summarized in Table~\ref{tbl:Parameters}.

For excitation and detection of the bulk acoustic waves we use two-color differential reflectivity pump-probe technique in the backscattering geometry. A dual oscillator system Gigajet 20/20c from Laser Quantum provides $\sim$100-fs pump and probe pulses at a repetition rate of about $f_0\approx1$~GHz, which are locked together at a repetition offset frequency of 20 kHz. An asynchronous optical sampling system (ASOPS) allows high-speed scanning over a nanosecond time delay between the pump and probe pulses~\cite{SubharmCombMembr2011}. 
The pump and probe wavelengths can be tuned independently within the range of $750-850$~nm (the corresponding photon energy range is $1.459-1.653$~eV).

For excitation and detection of the bulk acoustic waves we use two-color differential reflectivity pump-probe technique in the backscattering geometry. A dual oscillator system Gigajet 20/20c from Laser Quantum provides $\sim$100-fs pump and probe pulses at a repetition rate of about $f_0\approx1$~GHz, which are locked together at a repetition offset frequency of 20 kHz. An asynchronous optical sampling system (ASOPS) allows high-speed scanning over a nanosecond time delay between the pump and probe pulses~\cite{SubharmCombMembr2011}. 
The pump and probe wavelengths can be tuned independently within the range of $750-850$~nm (the corresponding photon energy range is $1.459-1.653$~eV).
Using a second-harmonic generation unit, the pump wavelength is set to 410~nm (photon energy 3.024~eV), which lies above the band gap of the studied semiconductors and ensures high absorption near the surface. Indeed, the absorption coefficient at the pump wavelength is $\sim 6 \times 10^{5}\,\mathrm{cm^{-1}}$ for GaAs~\cite{Sturge1962} and $\sim 2 \times 10^{5}\,\mathrm{cm^{-1}}$ for Cd$_{0.88}$Zn$_{0.12}$Te~\cite{OptConstCdZnTe}, corresponding to absorption lengths of only $\sim 17$ and $\sim 50\,\mathrm{nm}$, respectively.

The wavelength of the probe laser is varied in the transparency region for each material. A pulse shaper was used for spectral narrowing of the probe light to the spectral width of 1~nm (which corresponds to $1.7-2.2$~meV in the investigated spectral range). The spectrally narrow probe laser is essential for detection of sharp optical resonances resulting from multiple strain pulses. A microscope objective with $20\times$ magnification and a numerical aperture of 0.40 is used to focus the pump and probe beams at normal incidence onto the same spot on the sample surface, producing diameters of approximately 4 and $3\,\mu\mathrm{m}$, respectively. The pump energy density per pulse is $\sim 100\,\mu\mathrm{J}/\mathrm{cm}^2$, while the probe energy density is kept at $\sim 7\,\mu\mathrm{J}/\mathrm{cm}^2$. The probe light reflected from the sample is then collected with the same microscope objective, and the pump wavelength is filtered out by a long-pass spectral filter. The change in reflectivity as a function of the delay time between the pump and probe pulses $\Delta R(t)/R$ is detected using fast photodiode in combination with a digitizer card, providing overall resolution of better than 10~ps. All experiments are performed at low temperature of $T = 5\,\mathrm{K}$ using a helium flow cryostat (Oxford Microstat).

\section{Results and discussion}

\subsection{Generation and detection of the acoustic Bragg gratings in GaAs and CdZnTe}

\begin{figure}
  \centering
  \includegraphics[width=1\linewidth]{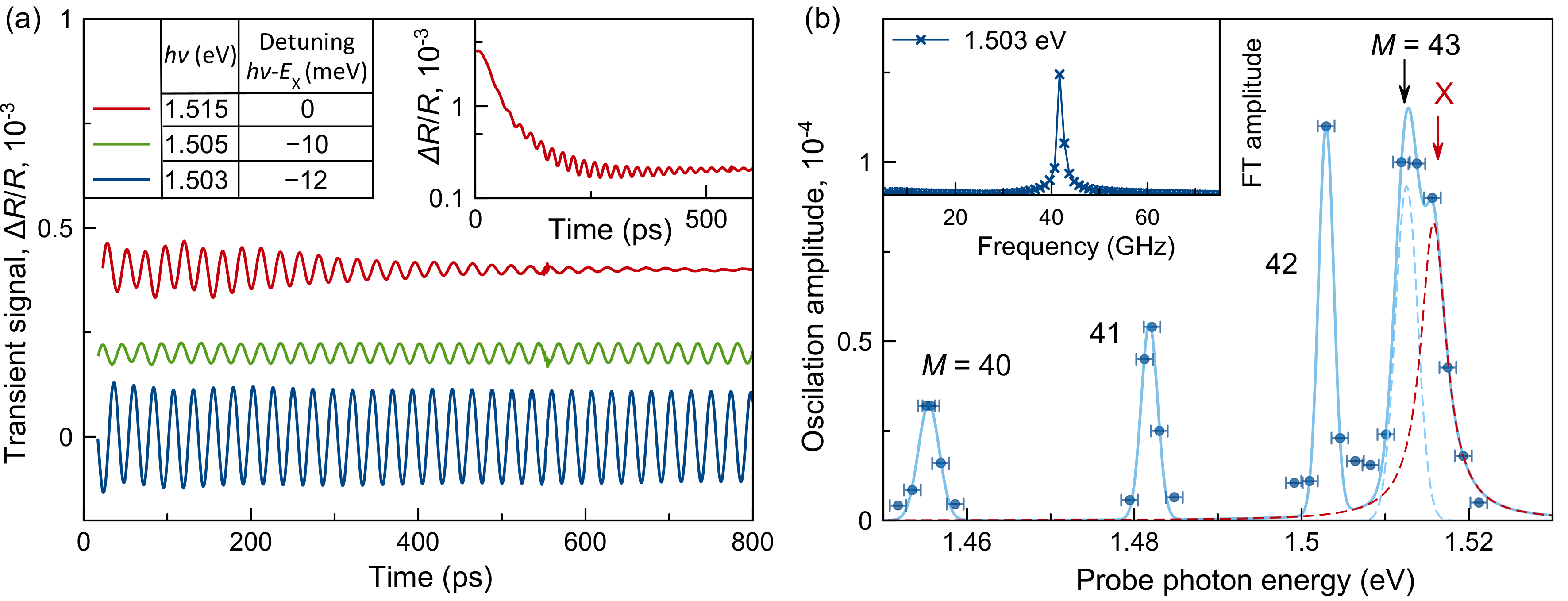}
  \caption{Results for GaAs. (a) Oscillatory component of the transient signals measured at different probe photon energies $h\nu$ (the exponential background has been subtracted, and the curves are vertically offset for clarity). The inset shows the raw signal at $h\nu=E_{\mathrm{X}}=1.515\, \mathrm{eV}$. (b) Spectral dependence of the oscillatory signal amplitude (blue dots) revealing peaks which correspond to the interference orders $M = 40-43$ and the exciton (X) line. Horizontal error bars indicate the probe spectral width. The solid blue curve shows a global fit given by the sum of Gaussian functions for the interference peaks and a Lorentzian function for the exciton. The dashed light-blue and red curves represent the corresponding individual Gaussian and Lorentzian contributions. The inset shows the FT spectrum of the oscillatory signal component measured at the probe photon energy $h\nu=1.503\, \mathrm{eV}$.}
  \label{fig:GaAs}
\end{figure}

Figure~\ref{fig:GaAs}(a) shows representative $\Delta R(t)/R$ transients measured for the [100] GaAs sample at different probe photon energies. A typical raw signal (see the inset in Fig.~\ref{fig:GaAs}(a)) consists of a slowly decaying background superimposed with fast oscillations. The slow decay originates from reflectivity changes associated with the recombination of charge carriers excited by the pump pulse and is subtracted for further data processing. 

The oscillatory component of the signal can be fitted by a damped sine wave. For the probe photon energy $h\nu = 1.515\, \mathrm{eV}$, which corresponds to the exciton resonance (upper red curve in Fig.~\ref{fig:GaAs}(a)), the damping is strong, with a characteristic decay time of $\sim 200\, \mathrm{ps}$, reflecting the large absorption coefficient at this photon energy. In contrast, transients recorded at  smaller photon energies, within the GaAs transparency range, show no appreciable decay as depicted by the lower green and blue curves in Fig.~\ref{fig:GaAs}(a).
A Fourier-transform (FT) spectrum of the oscillatory transient signal measured at $h\nu=1.503\, \mathrm{eV}$ (see inset in Fig.~\ref{fig:GaAs}(b)) shows a pronounced peak at the oscillation frequency $f_{\mathrm{ac}} = 42.1 \pm 0.7\,\mathrm{GHz}$, which corresponds to the longitudinal sound velocity of GaAs, $v_{\mathrm{ac}}=4.73\,\mathrm{km/s}$~\cite{Blakemore1982}, and its refractive index, $n \approx 3.66$~\cite{Sell1974}.Remarkably, variation of the probe photon energy by only $2\, \mathrm{meV}$ 
leads to a change in the oscillation amplitude exceeding one order of magnitude (see lower green and blue curves in Fig.~\ref{fig:GaAs}(a)). Figure~\ref{fig:GaAs}(b) shows the amplitude of the oscillatory component of the $\Delta R(t)/R$ signal plotted against the probe photon energy representing a series of sharp peaks -- in full agreement with the proposed concept.

For GaAs, in the studied range of probe wavelengths, we observe 4 narrow peaks with a full width at half maximum (FWHM) of approximately 2~meV (Fig.~\ref{fig:GaAs}(b)). The corresponding interference orders $M$ can be calculated from Eq.~(\ref{eq: resonance}), which results in $M=40-43$. 
The spectral dependence of the oscillation amplitude of the $\Delta R/R$  signal also shows a strong increase near the exciton resonance. On the one hand, as the photon energy of the probe approaches the band edge, the absorption coefficient increases substantially, reducing the penetration length of the probe light to $\sim 1$~$\mu$m at the exciton resonance. This means that at the exciton line, the amplification mechanism described above is no longer effective. Nevertheless, the strong dispersion of the refractive index at the exciton resonance leads to an exciton-induced enhancement of the $\Delta R/R$ signal: the elastic strain pulse modulates the exciton energy, so that  even a small deformation produces a noticeable change in reflectivity at the detection wavelength~\cite{AkimovUltrafast2006}. Indeed, an additional peak at around 1.515~eV -- corresponding to the exciton resonance at 5~K -- can be seen in the spectrum in Fig.~\ref{fig:GaAs}(b). This is consistent with the exciton-induced increase in the TDBS signal reported previously in Refs.~\cite{Scherbakov2022, 72ym-hc78}.

\begin{figure}
 \centering
  \includegraphics[width=1\linewidth]{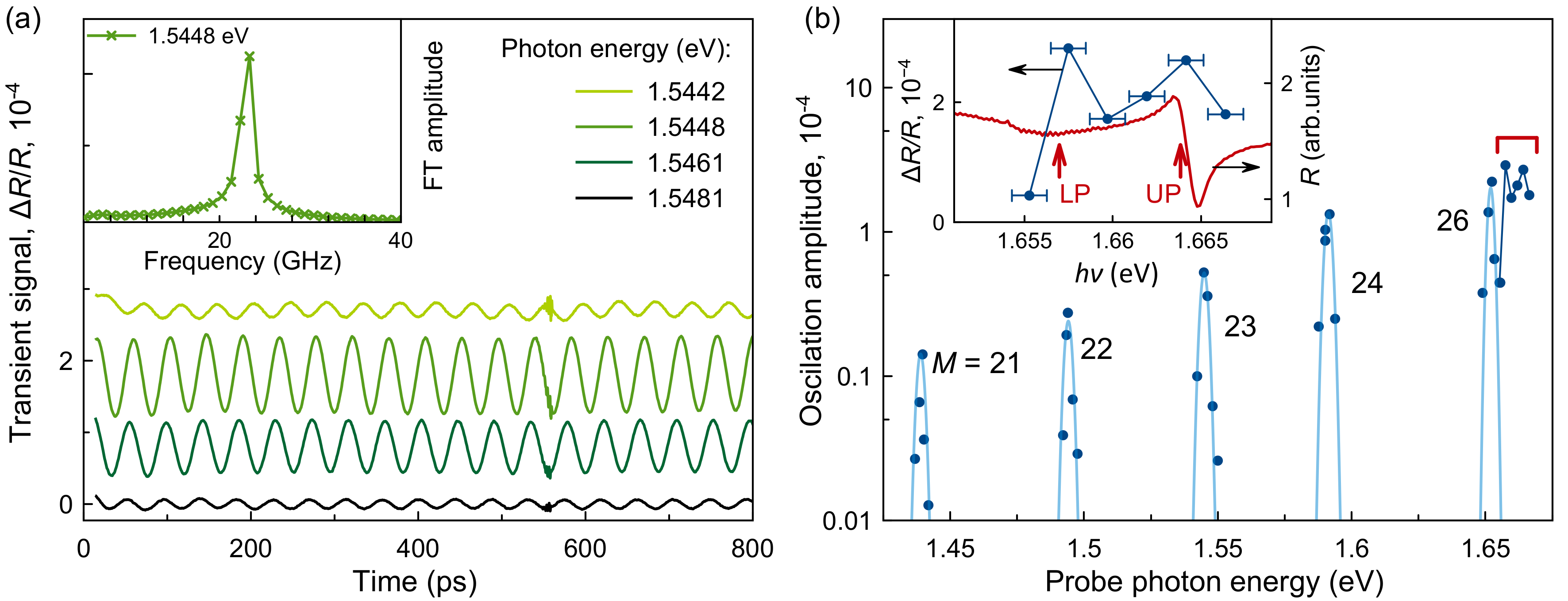}
  \caption{Results for (Cd,Zn)Te. (a) Variation of the oscillatory components of the transient signals 
  along the contour of the $M=23$ Bragg maximum (the exponential background has been subtracted, and the curves are vertically offset for clarity). Inset shows the FT spectrum of the oscillatory component of the signal measured at a probe photon energy $h\nu=1.5448\, \mathrm{eV}$.  (b) Spectral dependence of the oscillatory signal amplitude (blue dots) showing peaks which correspond to interference orders $M = 21-26$ and their fit by Gaussian functions (light-blue solid line).
  Inset shows enlarged spectral region indicated by the bracket in the main panel: the oscillatory signal amplitude (blue dots  connected by lines) and the reflectivity spectrum of the same sample (red solid line) adapted from Ref.~\cite{GoddeAkimov2010}. Horizontal error bars indicate the probe spectral width. Vertical arrows indicate the lower (LP) and upper (UP) polariton branches.}
  \label{fig:CdZnTe}
\end{figure}

A similar behavior is observed for (Cd,Zn)Te. Figure~\ref{fig:CdZnTe}(a) shows how the amplitude of the oscillations in the transient reflectivity signal, $\Delta R(t)/R$, varies with small changes in the probe photon energy. For (Cd,Zn)Te, the oscillation frequency is lower than in GaAs, reaching $f_{\mathrm{ac}} \approx 23\,\mathrm{GHz}$ at $h\nu = 1.5448\,\mathrm{eV}$, as evidenced by the FT spectrum shown in the inset. This value corresponds to the refractive index of (Cd,Zn)Te of $n \approx 2.91$~\cite{CdZnTe_refr} and the sound velocity $v_{\mathrm{ac}} = 3.15\,\mathrm{km/s}$, which is close to the longitudinal sound velocity of CdTe that can be calculated using the elastic moduli reported in Refs.~\cite{elasticCdTe1962, elasticCdTe2006, elasticCdTe2022}. Owing to these smaller values of $v_{\mathrm{ac}}$ and $n$, the Bragg condition is satisfied over a denser set of probe wavelengths, resulting in a larger number of well-resolved Bragg peaks. Accordingly, the spectral dependence of the oscillation amplitude exhibits distinct peaks with a FWHM of approximately $3\,\mathrm{meV}$ corresponding to the interference orders $M=21-26$ (Fig.~\ref{fig:CdZnTe}(b)).

Near the exciton resonance, $E_{\mathrm{X}}=1.664\, \mathrm{eV}$, two additional peaks at a distance of $\sim 10\, \mathrm{meV}$ are observed. It is well established that in bulk semiconductors the strong exciton-photon interaction results in the formation of exciton polaritons ~\cite{Polaritons1,Polaritons2, Polaritons3}. Indeed, the two features at photon energies of 1.664 and 1.657~eV previously observed in the reflectivity and PL spectra of the (Cd,Zn)Te sample were attributed to optical transitions involving the upper polariton (UP) and lower polariton (LP) branches, respectively ~\cite{GoddeAkimov2010}. The inset in Fig.~\ref{fig:CdZnTe}(b) presents a comparison of the amplitude of the TDBS signal with the reflectivity spectrum from Ref.~\cite{GoddeAkimov2010} showing pronounced UP- and LP-related peaks. The exciton-polariton resonance gives rise to the amplification of the TDBS signal because of the strong change in the refractive index near the resonance -- as described above for the exciton resonance in GaAs.

\subsection{Modeling of the Bragg-enhanced TDBS signal}

Conventional TDBS with a single strain pulse is typically based on heterodyne detection of the coherent acoustic waves~\cite{Ruello2015, Gusev2018}. Specifically, the weak probe field scattered by the acoustic waves and transmitted into air, $E_s$, interferes with the much stronger probe field reflected from the sample surface, $E_r$.

The normalized reflectivity change can be expressed as follows:
\begin{equation}
 \frac{\Delta R(t)}R = \frac{|E_{r}+E_{s}|^{2}-|E_{r}|^{2}}{|E_{r}|^{2}} \simeq \frac{2\operatorname{Re}\,(E_{r}^*E_{s})}{|E_{r}|^{2}} \equiv 2\operatorname{Re}\,S_0(t),
\end{equation}
where we neglect the smallest term $|E_r|^2$, which corresponds to the intensity of the scattered light.

Here $S_0(t) \propto \exp (i2\pi f_{\mathrm{ac}}t-t/\tau)$ is the conventional TDBS signal oscillating with the time delay $t$ between the pump and probe pulses. The decay time $\tau$ of the TDBS signal is determined by the optical absorption of the probe light ($\alpha$) and the acoustic attenuation coefficient ($\beta$), which describes damping of the strain pulse during propagation in the crystal. These contributions can be combined into an effective attenuation coefficient of the TDBS signal, $\gamma$: 
\begin{equation}
1/\tau = (\alpha + \beta)v_{\mathrm{ac}}\equiv \gamma v_{\mathrm{ac}}.
\label{eq:tau_gamma}
\end{equation}

In our experiments, the samples are excited by a pulsed laser with a high repetition rate ($\sim$1~GHz), resulting in the formation of a dynamic acoustic grating with a fixed period $L = v_{\mathrm{ac}}/f_0$. It propagates into the bulk of the sample at the sound velocity and periodically modulates the local refractive index. The probe beam penetrates the crystal and is weakly scattered by each of the strain pulses. In a backscattering geometry, the probe light scattered from successive pulses interferes. 
For simplicity, we consider the idealized infinite-coherence limit, i.e., an acoustic grating formed by an infinite number of strain pulses ($N \to \infty$). Convergence of the sum of scattered probe fields—forming a geometric series—is ensured by the effective attenuation coefficient. The full derivation is given in Supporting Information. The resulting detected signal is:

\begin{equation}
\begin{aligned}
&\frac{\Delta R(t)}R \simeq 2\operatorname{Re}\,S(t),
& S(t) = \frac{S_0(t)}{1-e^{-\gamma L}e^{i\varphi}},
\end{aligned}
\label{eq:signal_S}
\end{equation}
where $\varphi = 
4\pi n L \nu/c$ denotes the phase that governs the interference of the scattered probe light with optical frequency $\nu$. 
Equation~(\ref{eq:signal_S}) has the same functional form as the intracavity field (cavity response function) of a Fabry–Perot interferometer~\cite{BornWolf,Lasers1986}. Its squared modulus yields the Airy line shape characteristic of multi-beam interference.

When the Bragg condition (Eq.~(\ref{eq: resonance})) is satisfied, $\varphi (h\nu_M) = \varphi_{M} = 2\pi M$ and constructive interference of the probe light scattered from successive pulses gives rise to a series of resonant peaks in the spectral dependence of the signal amplitude at the discrete probe photon energies $h\nu_M=hc/\lambda_{M}$ (with $h$ being the Planck constant). In the limit of weak attenuation ($\gamma L \ll 1$), the signal is resonantly enhanced by a factor $|S/S_0| =1/(1-e^{-\gamma L}) \approx (\gamma L)^{-1} \gg~1$.

According to Eq.~(\ref{eq:signal_S}), the spectral dependence of the peak amplitude $|S(h\nu_M)|$ is governed by both the single-pulse TDBS response $|S_0(h\nu)|$ and the effective attenuation coefficient $\gamma(h\nu)$, which act in opposite ways. With increasing probe photon energy, the strain-induced modulation of the refractive-index $\Delta n$ increases due to the proximity of the optical resonances (either exciton or exciton-polariton). It should be noted that the spectral dependence of $\Delta n$ is determined not only by the exciton resonance at the fundamental band-gap edge at the center of the Brillouin zone, but also by higher-lying transitions with larger oscillator strengths~\cite{Furdyna2004,Godde2013}.
This enhancement increases the reflection coefficient of the propagating strain pulse, $r \approx \Delta n/(2n)$, and therefore the single-pulse TDBS amplitude $|S_0| \propto |r|$ (see Supporting Information).
At the same time, the absorption coefficient $\alpha$ increases, leading to a larger effective attenuation $\gamma$, which partially offsets the growth of the signal amplitude $|S(h\nu_M)|$ toward higher photon energies. Overall, these considerations qualitatively explain the observed increase in the peak amplitude with photon energy, as shown in Figs.~\ref{fig:GaAs}(b) and ~\ref{fig:CdZnTe}(b).

Close to the signal maxima, when $\lvert \Delta h\nu \rvert = \lvert\nobreak h\nu-h\nu_{M} \nobreak\rvert \ll h\nu_M$, the spectral shape of the peaks can be described by a square root of the Lorentzian function, while the phase of the signal, $\delta$, changes sign along the spectral contour:

\begin{equation}
\begin{aligned}
&|S(h\nu)| \propto \frac{1}{\sqrt{1 + (h\nu-h\nu_{M})^2/\Gamma^2}}, \\
&\delta(h\nu) = \arctan\!\left( \frac{h\nu - h\nu_{M}}{\Gamma} \right),
\end{aligned}
\end{equation}
where $2\Gamma = hc\gamma (2 \pi n)^{-1}$ is the characteristic linewidth (a full derivation is presented in the Supporting Information).

For both GaAs and (Cd,Zn)Te, the FWHM of the Bragg peaks is approximately equal to the probe spectral width ($\sim 1\,\mathrm{nm}$), yielding attenuation coefficients $\gamma \approx 200$ and $100\,\mathrm{cm}^{-1}$ and corresponding resonant enhancement factors $(\gamma L)^{-1}$ of $\sim 10$ and $\sim 30$, respectively. The latter values are consistent with the ratio between the maximum and minimum oscillation amplitudes of the $dR/R$ signal for individual Bragg peaks. However, the actual attenuation $\gamma=\alpha+\beta$, and thus the intrinsic peak width, may be significantly smaller, implying correspondingly larger enhancement factors. For example, in single-crystal GaAs at $T=5\,\mathrm{K}$, the optical absorption coefficient for $h\nu<1.5\,\mathrm{eV}$ is $\alpha<10\,\mathrm{cm}^{-1}$~\cite{Sturge1962}, while the acoustic attenuation $\beta$ is even smaller~\cite{phn_atten1,phn_atten2}, suggesting a potential enhancement factor of up to $(\gamma L)^{-1}\sim 200$. 

In practice, however, the enhancement is limited by the finite number $N$ of acoustic ``layers'' (strain pulses) within the sample thickness $d$, yielding a maximum $|S/S_0|\sim N=d/L\approx 100$ and 150 for the present GaAs and (Cd,Zn)Te samples ($d\approx 500\,\mu\mathrm{m}$), respectively. Under the present experimental conditions, a more restrictive bound is imposed by the finite probe bandwidth ($\sim 1\,\mathrm{nm}$), or equivalently the pulse duration $\Delta t\sim 1\,\mathrm{ps}$. For interference-based enhancement, the probe fields reflected from the surface and scattered from the $N$th acoustic layer must temporally overlap. This condition, governed by the autocorrelation of two pulses separated by the round-trip delay $T=2NnL/c$, requires $T\leq 2\Delta t$, which limits the number of coherently contributing layers to $N\sim 20$ and 30 for GaAs and (Cd,Zn)Te, respectively. These values are consistent with the experimentally extracted enhancement factors, indicating that the observed Bragg peaks are resolution-limited and can be well described by Gaussian functions (see Figs.~\ref{fig:GaAs}(b) and ~\ref{fig:CdZnTe}(b)). Additional limitation of the resonant enhancement factor arises from the generation of the acoustic Bragg grating itself: in practice, the acoustic wavefront is not perfectly planar but exhibits finite curvature due to excitation by a focused laser beam. This effect, together with the limitation imposed by the finite probe duration, can be incorporated phenomenologically as additional damping contribution in Eq.~(\ref{eq:tau_gamma}), leading to a reduction of the decay time $\tau$ and an increase in the effective attenuation $\gamma$ of the TDBS signal $S_0(t)$.

Resolving the intrinsic peak shape and achieving the maximum enhancement require further narrowing of the probe spectrum, i.e., increasing the probe pulse duration. In this case, probe pulses scattered from deeper lying regions of the acoustic grating contribute coherently to the interference. However, spectral narrowing is constrained by the temporal resolution required for the TDBS measurements. As discussed in Ref.~\cite{Nelson1987}, where frequency- and time-domain light-scattering techniques were compared, the probe pulse duration $\Delta t$ must remain sufficiently short to resolve oscillations at the acoustic frequency. Specifically, for the present conditions, $\Delta t \leq 0.2/f_{\mathrm{ac}}\sim 5-10\, \mathrm{ps}$ is required, which sets a lower bound on the probe spectral width of $\sim 0.1-0.2\, \mathrm{nm}$ in the investigated spectral range. These parameters, however, are more than sufficient to access the entire acoustic grating across the whole thickness of both samples.

Additionally, following Refs.~\cite{Nelson2011, Aleman2020, SciRep2018, SubharmCombMembr2011}, the formation of the acoustic grating can be analyzed in the frequency domain as frequency-selective excitation of coherent phonons. Under single-pulse excitation, the strain spectrum is broadband, extending up to a characteristic frequency $f\sim v_{\mathrm{ac}}\alpha_{\mathrm{pump}}$, where $\alpha_{\mathrm{pump}}$ is the absorption coefficient of the pump light. The acoustic frequency detected in the experiment, $f_{\mathrm{ac}}=2v_{\mathrm{ac}}n\nu/c$, is determined by the selection rules for Brillouin light scattering for a given geometry set by the heterodyne detection. Using the parameters listed in Table~\ref{tbl:Parameters}, we estimate the characteristic bandwidth for single-pulse excitation to be $\sim 300$ and $\sim 70\, \mathrm{GHz}$ for GaAs and (Cd,Zn)Te, respectively. 

In contrast, excitation by a periodic train of pump pulses leads to a preferential population of acoustic phonon states with frequencies that are integer multiples of the excitation repetition rate~\cite{Aleman2020}: $f_{\mathrm{ac}}^M=Mf_0$. These frequencies are selectively detected at probe photon energies $h\nu_M$ satisfying the Bragg condition, $Mf_0=2v_{\mathrm{ac}}n\nu_M/c$, and thus corresponding to maxima of the TDBS signal.
Increasing the number $N$  of pulses in the sequence (or, equivalently, the number of ``layers'' in the acoustic Bragg grating) leads to a progressive narrowing of the spectral peaks in the phonon spectrum approaching a quasi-monochromatic response in the limit $N \to \infty$~\cite{SciRep2018,Yan1991theory}. For the present experimental conditions ($f_0 = 1\,\mathrm{GHz}$ and $N \sim 10$–$30$), the expected linewidth of the peak in the acoustic phonon spectrum is $\Delta f \sim f_0/N \sim 30 - 100\, \mathrm{MHz}$. However, the resolution of the FT spectra (see insets in Figs.~\ref{fig:GaAs}(b) and ~\ref{fig:CdZnTe}(a)) is limited to $\sim 1\,\mathrm{GHz}$ by the finite observation time window ($\sim 1\, \mathrm{ns}$), precluding direct observation of the actual narrowing of the acoustic spectrum.

\section{Conclusion}

We have demonstrated a new approach to ultrafast acoustic interferometry in semiconductors that employs high-repetition-rate ($\sim 1$~GHz) femtosecond laser pulses with photon energies above the semiconductor band gap. This method generates a dynamic one-dimensional acoustic grating that propagates into the bulk at the sound velocity. The approach has been validated on prototypical semiconductor systems, namely bulk crystalline [100] GaAs and [100] (Cd,Zn)Te. We have also developed a quantitative description of the underlying interference mechanism and its spectral selectivity.

In terms of phonon generation, the proposed approach enables frequency-selective excitation of coherent phonons, similar to Refs.~\cite{Nelson2011, SciRep2018, SubharmCombMembr2011, Aleman2020}. In contrast to the broadband acoustic phonon spectrum produced under single-pulse excitation, which extends to hundreds of GHz, excitation by a periodic pulse train with a 1~GHz repetition rate leads to a preferential population of acoustic phonon states at frequencies that are integer multiples of the repetition rate. This results in discrete peaks in the phonon spectrum with an estimated linewidth of $\Delta f \sim 30-100\, \mathrm{MHz}$.

In terms of detection, the proposed scheme provides enhanced sensitivity to weak signals. Bragg reflection of the incident probe light from the propagating acoustic grating leads to resonant enhancement of the TDBS signal amplitude at discrete probe wavelengths satisfying the Bragg condition. The enhancement factor relative to conventional single-pulse TDBS is governed by the attenuation of the coherent acoustic phonon wave packet and the optical absorption of the probe light over one grating period (i.e., the spacing between successive strain pulses). In our experiments, a $10-30$-fold enhancement has been achieved for GaAs and (Cd,Zn)Te, determined by the finite spectral width of the probe pulse ($\sim 1\, \mathrm{nm}$).  In the limit of weak attenuation, further spectral narrowing of the probe should result in significantly larger enhancements of the order $\sim 100-150$, set by the number $N$ of ``layers'' (strain pulses) of the acoustic grating within the sample and ultimately limited by material parameters and sample thickness.

An additional enhancement of the TDBS signal has been observed in the vicinity of the exciton and exciton–polariton resonances in GaAs and (Cd,Zn)Te, respectively. This behavior is consistent with the strong dispersion of the refractive index near these resonances and is in line with the previously reported exciton-induced enhancement of the TDBS signal in the conventional single-pulse regime~\cite{Scherbakov2022, 72ym-hc78}.

\bibliography{refs2}

\newpage

\section{\textbf{Supporting Information} }

\subsection{Optical response in TDBS from an acoustic Bragg grating}

\begin{figure}[ht]
  \centering
  \includegraphics[width=0.6\linewidth]{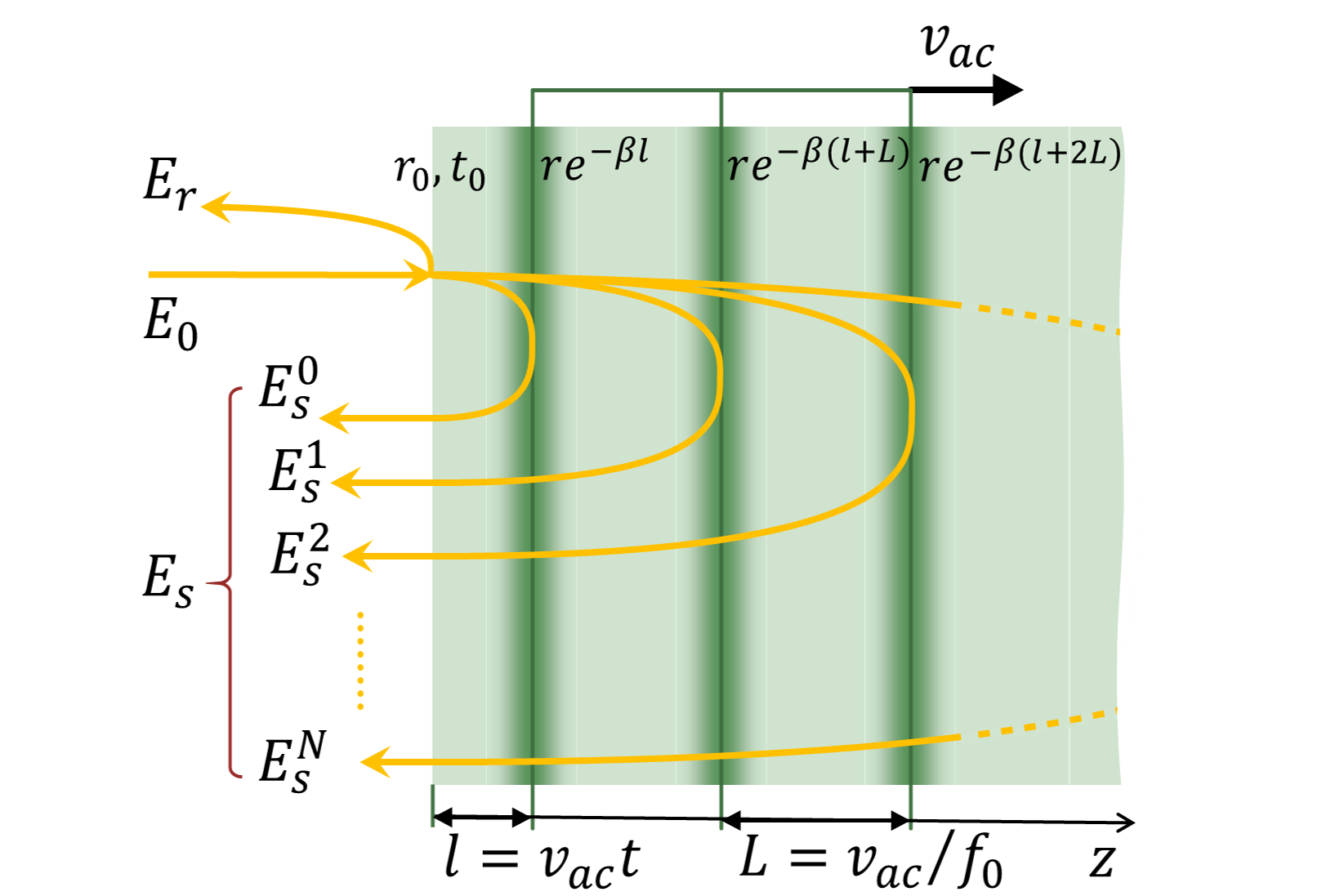}
  \caption{Schematic illustration of Bragg reflection from a propagating acoustic grating, giving rise to multi-beam interference of the probe light scattered by the different acoustic pulses.}
  \label{fig:appendix}
\end{figure}

We consider a probe beam incident normally onto the sample surface (Fig.~\ref{fig:appendix}) with electric field amplitude $E_0$, wave vector $k = 2\pi/\lambda = 2\pi \nu /c$, vacuum wavelength $\lambda$, and frequency $\nu$ (here $c$ is the speed of light in vacuum). The amplitude of the wave reflected from the sample surface is $E_r = r_0 E_0$, where $r_0$ is the surface reflection coefficient. Absorption of a femtosecond pump pulse in the near-surface region launches a coherent acoustic strain pulse. This pulse propagates into the bulk with the sound velocity $v_{\mathrm{ac}}$ and modulates the local refractive index: $n \rightarrow (n+\Delta n)$, giving rise to the reflection coefficient of the acoustic pulse of $r \approx \Delta n/(2n) \ll 1$, when modulation is weak ($\Delta n~\ll~n$). A delayed probe pulse (delay time $t$) penetrates into the crystal and is weakly scattered by the acostic pulse. In the backscattering geometry, the amplitude of the light field scattered by a single acoustic pulse can be written as: 
\begin{equation}
\begin{aligned}
E_s^0(t) = 
r t_0^2 E_0 \exp{(i 2\pi f_{\text{ac}} t -\gamma v_{\text{ac}} t)}, 
\end{aligned}
\end{equation}
where $t_0$ is the transmission coefficient at the sample surface, and we introduce the acoustic frequency $f_{\mathrm{ac}}=2v_{\mathrm{ac}}n/\lambda=2v_{\mathrm{ac}}n\nu/c$ and effective attenuation coefficient $\gamma = \alpha + \beta$, where $\alpha$ is the optical absorption coefficient and $\beta$ is an acoustic attenuation coefficient, which describes damping of the strain pulse during its propagation in the crystal.

In the standard single-pulse TDBS configuration the probe light is scattered by a single acoustic pulse, $E_s(t)=E_s^0(t)$, and the detected reflectivity change is given by

\begin{equation}
\begin{aligned}
\frac{\Delta R(t)}{R} = \frac{\left|E_r + E_s\right|^2 - \left|E_r\right|^2}{\left|E_r\right|^2}
\simeq \frac{2 \,\mathrm{Re}\!\left(E_r^{*} E_s^0\right)}{\left|E_r\right|^2} = 2\operatorname{Re}\!S_0(t),
\end{aligned}
\end{equation}
in which 
\begin{equation}
\begin{aligned}
S_0(t) = t_0^2 r/r_0 \exp{(i 2\pi f_{\text{ac}} t -\gamma v_{\text{ac}} t)} \equiv |S_0(t)| e^{i\Psi(t)}.
\end{aligned}
\label{eq:S0}
\end{equation}

We now consider excitation of a periodic sequence of acoustic pulses by a pump pulse train with a repetition rate $f_0$. The spatial period of the resulting acoustic grating is $L = v_{\mathrm{ac}}/f_{0}$. Considering weak modulation of the local refractive index of material by the acoustic pulse ($\Delta n~\ll~n$ and thus $r \ll 1$), we neglect multiple reflections between the ``layers'' of the grating. We also consider the transmission coefficient through each ``layer'' to be $\approx 1$.
Then the field scattered from the $N$-th acoustic pulse can be written as:
\begin{equation}
\begin{aligned}
E_s^N(t) & = r t_0^2 E_0 \exp{[i\left(2\pi f_{\text{ac}} t + 2 k n N L\right)]} \\
&\times \exp\!\left[-\gamma \left(v_{\text{ac}} t + N L\right)\right]
\end{aligned}
\end{equation}

The total field of the scattered wave, $E_s(t) = \sum E_s^{N}(t)$, in the limit $N \to \infty$ is given by a sum of the convergent geometric series:

\begin{equation}
\begin{aligned}
E_s(t) &= 
r t_0^{2} E_0 \exp\!\left(i 2\pi f_{\mathrm{ac}} t - \gamma v_{\mathrm{ac}} t\right) \sum_{N=0}^{\infty}
e^{N(i\varphi - \gamma L)}=\\
& = \frac{
r t_0^{2} E_0 \,
\exp\!\left(i 2\pi f_{\mathrm{ac}} t - \gamma v_{\mathrm{ac}} t\right)}{1 - \exp\!\left(i\varphi - \gamma L\right)
},
\end{aligned}
\end{equation}
where $\varphi = 2 k n L$. Thus, the detected signal becomes:
\begin{equation}
\begin{aligned}
&\frac{\Delta R(t)}R \simeq 2\operatorname{Re}\,S(t),
& S(t) = \frac{S_0(t)}{1-e^{-\gamma L}e^{i\varphi}},
\end{aligned}
\label{eq:Signal_S_app}
\end{equation}

Under the Bragg condition $Mc/\nu_M = 2 n L$ (where $M$ is integer) the phase of the signal is $\varphi_M = 2\pi M$ and the constructive interference leads to a maximum signal amplitude:
\begin{equation}
|S(\nu_M)| = \frac{|S_0|}{1 - e^{-\gamma L}}.
\end{equation}

For weak attenuation ($\gamma L \ll 1$), we obtain a large amplitude enhancement compared to the single-pulse TDBS: $|S(h\nu_M)|/|S_0| \approx (\gamma L)^{-1} \gg 1$.

Close to a resonance, when $|\Delta \nu| = |\nu_M - \nu| \ll \nu_M$, the phase is $\varphi = \varphi - \varphi_M$ with $\Delta \varphi = 2\pi M \Delta \nu/\nu_M \ll 1$.
Expanding the denominator in Eq.~(\ref{eq:Signal_S_app}) and supposing $\gamma L \ll 1$, we obtain:

\begin{equation}
S(t) \approx \frac{S_0(t)}{\gamma L + i\Delta\varphi} = \frac{|S_0|}{\gamma L} \frac{e^{i(\Psi(t) + \delta)}}
{\sqrt{1 + (\Delta h\nu / \Gamma)^2}},
\label{eq:S_Form}
\end{equation}
where $\Gamma = hc\gamma(4\pi n)^{-1}$ is the linewidth, while the phase shift is given by $\delta =-\arctan(\Delta h\nu/\Gamma)$ .

\end{document}